\documentclass[aps,pra,twocolumn,nofootinbib]{revtex4-1}
\usepackage{epsfig}
\usepackage[colorlinks,linkcolor=blue,anchorcolor=blue,citecolor=blue,urlcolor=blue,breaklinks=true]{hyperref}
\usepackage{graphics}
\usepackage{color}

\usepackage{epstopdf}
\usepackage{amsmath}
\usepackage{ulem}

\begin{document}
\author{Guo-Hua Liang$^{1}$}
\author{Yong-Long Wang$^{1,2,}$}~\email[]{Email: wangyonglong@lyu.edu.cn}
\author{Long Du$^{1}$}
\author{Hua Jiang$^{1,2}$}
\author{Guang-Zhen Kang$^{1}$}
\author{Hong-Shi Zong$^{1,3,4,}$}~\email[]{Email: zonghs@nju.edu.cn}

\address{$^{1}$Department of Physics, Nanjing University, Nanjing 210093, China}
\address{$^{2}$School of Science and Institute of Condensed Matter Physics, Linyi University, Linyi 276005, P. R. China}
\address{$^{3}$Joint Center for Particle, Nuclear Physics and Cosmology, Nanjing 210093, China}
\address{$^{4}$State Key Laboratory of Theoretical Physics, Institute of Theoretical Physics, CAS, Beijing, 100190, China}

\title{Coherent electron transport in a helical nanotube}
\begin{abstract}
The quantum dynamics of carriers bound to helical tube surfaces is investigated in a thin-layer quantization scheme. By numerically solving the open-boundary Schr$\ddot{\rm o}$dinger equation in curvilinear coordinates, geometric effect on the coherent transmission spectra is analysed in the case of single propagating mode as well as multimode. It is shown that, the coiling endows the helical nanotube with different transport properties from a bent cylindrical surface. Fano resonance appears as a purely geometric effect in the conductance, the corresponding energy of quasibound state is obviously influenced by the torsion and length of the nanotube. We also find new plateaus in the conductance. The transport of double-degenerate mode in this geometry is reminiscent of the Zeeman coupling between the magnetic field and spin angular momentum in quasi-one-dimensional structure.

\end{abstract}

%\pacs{73.22.Gk, 73.23.Ad, 73.63.Fg}

\maketitle

\section{INTRODUCTION}
The realization of growing quasi-two-dimensional surfaces of arbitrary shape in nanoscale helps people find new physical effects which are originated from the topology. Many intriguing phenomena associated with the surface curvatures, such as electron localization[1-3], Aharonov-Bohm oscillations[4,5] and anisotropic magnetoresistance[6], have been investigated. Briefly speaking, in both theoretical and experimental fields scientists have accomplished essential developments for the curved two-dimensional (2D) systems.

To describe a particle confined to a curved surface, there is a triumphant approach that is introduced by Jensen and Koppe[7] and da Costa[8] (JKC). In this approach a confining potential is introduced to squeeze[9] the particle on curved surface. The introduced potential gives rise to that the quantum excitation energies in the direction normal to the surface are substantially larger than those in the tangential directions. Hence one can reasonably neglect the particle motion in the normal direction, and focus on the effective and dimensionally reduced equation.
It is a great achievement to the JKC method that a curvature-induced potential appears in the effective 2D equation. The induced potential is the well-known geometric potential. The JKC approach has been successfully applied to many nanostructures with different geometries, such as rolled-up nanotubes[6,10], M$\ddot{\rm o}$bius stripes[3,11] and helicoidal ribbon[12]. And the method is also proved by experimental results[13-15], such as the geometric effects on electron states[15], on proximity effects[16], and on the transport in photonic topological crystals[17].

In past decades, various interesting properties in carbon nanotubes have been widely and deeply studied, such as quantum transport and conductance[18-24], size effects[25-27]. Quantum transmission is a natural property of nanostructure devices, in which the topological effect is considerable. Recently, the geometric effects on the coherent electron transport have been investigated in bent cylindrical surfaces[28], and in the surface of a truncated cone[29]. Additionally, the curvature effects on vitrification behavior has been discussed for polymer nanotubes[30]. In the nanotubes the geometrical curvature plays an important role to influence their quantum properties. At the same time, in twisted nanoscale systems some quantum properties and phenomena have been studied, such as bound states[31-33], coherent electron transport[32,34], spin-orbit coupled electron[35]. In terms of those investigations, one can realize that the torsion-induced effect is significant to the quantum properties of the twisted systems. Consequently, in the present study we will investigate the coherent electron transport in helically coiled nanotubes (hereafter referred to as helical nanotubes) with finite length.

In this work, we treat the electron states in the effective mass approximation, which is valid for the conventional semiconducting nanotubes. The ballistic 1D transport in nanotubes has been demonstrated by several experiments[36-38]. In the case of semiconducting helical nanotubes, by taking into account the local change of electronic property[39,40] induced by geometric deformation, the envelope-function approach can still be used. Electron localization caused by the mixing of $\sigma$ and $\pi$ states are presented by the effective geometric potential in this approach.  We will employ quantum transmitting boundary method (QTBM)[41] to numerically solve the transmission probability. This method is capable of solving open-boundary transmission problems for arbitrary internal geometries, since it can be generalized to include the metric tensor of the system[28,32].

In the calculational procedure, it is treated as that the two components of effective mass tensor[42,43] in two directions on the surface of 2D nanotubes are equal.
To avoid misunderstanding, we stress that in our analysis, only the geometric chirality associated with torsion is considered and discussed, the effect of the chirality of atomic structure is ignored.

This work is structured as follows. In Sec. II, we outline the mathematical description of an electron confined to the surface of a helical tube, and analyse the geometric potential and modes in leads. In Sec. III, we numerically calculate the transmission probability in helical nanotubes and discuss the relationship between the transport and the symmetries in the helical system. In Sec. IV, the conductance at zero temperature is presented. Finally, in Sec. V, we have a brief summary.

\section{MODEL}

\subsection{Quantum dynamics of a particle constrained on a helical tube surface}
One can construct a helical tube (as shown in Fig. 1) by moving a disk with radius $\rho_0$ along a helical line parametrized as $\textbf{x}(s)$. To describe this geometry we introduce the Frenet frame vectors $\textbf{t}$, $\textbf{n}$ and $\textbf{b}$ which satisfy
\begin{equation}
\left(
\begin{array}{ccc}
\dot{\textbf{t}}\\
\dot{\textbf{n}}\\
\dot{\textbf{b}}
\end{array}
\right)=\left(
\begin{array}{ccc}
0&\kappa(s)&0\\
-\kappa(s)&0&\tau(s)\\
0&-\tau(s)&0
\end{array}
\right)\left(
\begin{array}{ccc}
\textbf{t}\\
\textbf{n}\\
\textbf{b}
\end{array}
\right),
\end{equation}
where $\textbf{t}$, $\textbf{n}$ and $\textbf{b}$ are the unit tangent vector, normal vector and binormal vector of $\textbf{x(s)}$, respectively, the dot denotes derivative with respect to the natural parameter $s$, and $\kappa(s)$ and $\tau(s)$ are the curvature and torsion of $\textbf{x}(s)$, respectively. During the disk moving along $\textbf{x}(s)$, the disk is always orthogonal to $\textbf{t}$, on the disk plane $\textbf{n}$ and $\textbf{b}$ shift due to $\tau(s)$. It is convenient to define two new vectors
\begin{equation}
\textbf{N}={\rm {cos}}\theta(s)\textbf{n}+{\rm {sin}}\theta(s)\textbf{b},
\end{equation}
\begin{equation}
\textbf{B}=-{\rm {sin}}\theta(s)\textbf{n}+{\rm {cos}}\theta(s)\textbf{b},
\end{equation}
where the angle $\theta(s)=-\int_{s_0}^s\tau(s^\prime)ds^\prime$.

\begin{figure}
\includegraphics[width=0.3\textwidth]{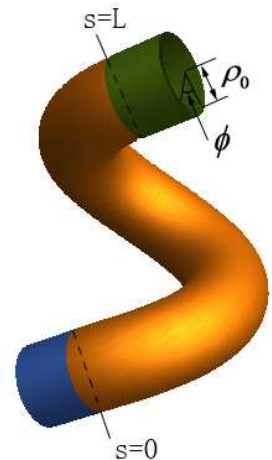}
\caption{ (Color online) Surface of a helical tube with two straight cylinders at the two ends. The geometry is parametrized by $s$ and $\phi$.}\label{helical}
\end{figure}

In this new frame, $\textbf{N}$ and $\textbf{B}$ are fixed on the disk. The relation Eq.(1) becomes
\begin{equation}
\left(
\begin{array}{ccc}
\dot{\textbf{t}}\\
\dot{\textbf{N}}\\
\dot{\textbf{B}}
\end{array}
\right)=\left(
\begin{array}{ccc}
0&\xi(s)&-\eta(s)\\
-\xi(s)&0&0\\
\eta(s)&0&0
\end{array}
\right)\left(
\begin{array}{ccc}
\textbf{t}\\
\textbf{N}\\
\textbf{B}
\end{array}
\right),
\end{equation}
where $\xi(s)=\kappa(s){\rm cos}\theta(s)$, $\eta(s)=\kappa(s){\rm sin}\theta(s)$.
Consequently, points on the tube surface can be parametrized as
\begin{equation}
\textbf{R}(s,\phi)=\textbf{x}(s)-\rho_0[{\rm sin}(\phi)\textbf{B}+{\rm cos}(\phi)\textbf{N}],
\end{equation}
where $\phi$ is the angular position of the point on the edge of the disk.

It's straightforward now to get the metric tensor of the tube surface by using the definition $g_{ij}=\frac{\partial\textbf{R}}{\partial q^i}\cdot\frac{\partial\textbf{R}}{\partial q^j}$ and the relation Eq.(4),
\begin{equation}
g_{ij}=\left(
\begin{array}{ccc}
w^2&0\\
0&\rho_0^2
\end{array} \right)   \qquad i,j=1,2,
\end{equation}
where $w=1+\rho_0\kappa(s){\rm cos}[\phi+\theta(s)]$ and $q^1$, $q^2$ refer to $s$ and $\phi$, respectively.
Besides, the Weingarten curvature matrix is also obtained
\begin{equation}
\alpha_{ij}=\left(
\begin{array}{ccc}
\frac{\kappa(s){\rm cos}(\phi+\theta(s))}{w}&0 \\
0&\frac{1}{\rho_0}
\end{array}
\right).
\end{equation}
By means of the matrix above, it is easy to obtain the mean curvature $M=\frac{1}{2}{\rm tr}(\alpha)$ and Gaussian curvature $K={\rm det}(\alpha)$.

Following da Costa[8] the well-known geometric potential is
\begin{equation} \label{vg}
V_g=-\frac{\hbar^2}{2\mu}(M^2-K)=-\frac{\hbar^2}{8\mu\rho_0^2w^2},
\end{equation}
where $\mu$ is the effective mass of carrier. For comparison with Ref.[28], $\mu=0.173m_e$, with $m_e$ the free electron mass.

In the effective-mass approximation, the envelop function associated to the energy $E$ is described by the time independent Schrodinger equation
\begin{equation} \label{helieq}
E\psi=-\frac{\hbar^2}{2\mu}\frac{1}{\sqrt{g}}\partial_i(\sqrt{g}g^{ij}\partial_j\psi)+V_g\psi ,
\end{equation}
where $\psi$ is a wave function and the repeated index summation convention is used. In our case, $\kappa$ and $\tau$ are supposed as constants, the corresponding Hamiltonian in the curvilinear coordinate system ($s,\phi$) can be explicitly expressed as
\begin{equation} \label{Hamiltonian}
\begin{aligned}
    \mathcal{H}_h=&-\frac{\hbar^2}{2\mu}\left[ \right.\frac{\partial_s^2}{w^2}-\frac{\rho_0\kappa\tau {\rm sin}(\phi-\tau s)\partial_s}{w^3}+\frac{\partial_\phi^2}{\rho_0^2}\\
&-\frac{\kappa {\rm sin}(\phi-\tau s)\partial_\phi}{\rho_0 w}\left. \right]
-\frac{\hbar^2}{8\mu\rho_0^2w^2}.
\end{aligned}
\end{equation}

\subsection{Geometric potential and modes in leads}
The surface of a helical nanotube $S_h$ together with two short straight cylinders (as leads) connected at the two ends is shown in Fig. ~\ref{helical}. As expected, when $\tau\rightarrow0$, the Hamiltonian Eq.(\ref{Hamiltonian}) is to describe the dynamics on a bent cylindrical surface[28] (here the tube length is assumed less than $2\pi/\kappa$, otherwise the surface will form a torus), furthermore, if $\kappa\rightarrow0$, the Hamiltonian becomes that for a straight cylindrical surface. It is worthwhile to notice that the longitudinal curvature of the surface is discontinuous at both the connections between $S_h$ and two straight cylindrical leads. However, it is easy to prove that wave function $\psi$ and its first-order derivative are continuous at the two connections. For example, at $s=0$ in Fig. 1, doing the integration $\int_{-\epsilon}^{\epsilon}w ds$ on both sides of Eq. (\ref{helieq}), and letting $\epsilon\rightarrow0$, we will obtain $\frac{d\psi}{ds}|_{s=0+}=\frac{d\psi}{ds}|_{s=0-}$. This gives the system an open boundary condition.

\begin{figure}
  \centering
  \includegraphics[width=0.47\textwidth]{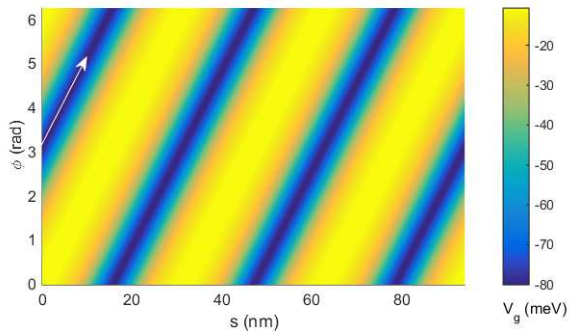}
  \caption{(Color online) The geometric potential of $S_h$ as a function of the curvilinear coordinates $s$ and $\phi$, with $\kappa$=0.3nm$^{-1}$, $\rho_0$=1.55nm, $\tau$=0.2nm$^{-1}$. Along the direction of the arrows the geometric potential has a constant value.}\label{potential}
\end{figure}

The geometric potential $V_g(s,\phi)$ is described in Fig. \ref{potential}.
It shows that the geometric potential $V_g(s,\phi)$ is periodic in both $\phi$ and $s$ directions. However, along the direction  $\phi=\tau s$ the potential has a constant value.
%The direction of $\phi$ is labeled by arrows as shown in Fig.2 (a) and Fig.2 (b). The slope of the arrows are equal to $\tau$.
The positions where maximums appear correspond to inner points on $S_h$, contrarily the minimums correspond to outer points. In compare to the constant potential in the leads $V_g=-\hbar^2/(8\mu\rho_0^2)$, the geometric potential on the outer part of $S_h$ provides an attractive effect for carriers, the inner part gives a repulsive effect.
%On the outer points the differences between two principle curvatures are maximum.

The chirality of the helical axis line $\textbf{x}(s)$ is determined by the sign of $\tau$. For the sake of simplicity, we  also define the chirality of the helical tube by the sign of the torsion $\tau$ of its axis. In geometric potential, the chirality is seen from the phase term $\phi-\tau s$ in the expression (\ref{vg}). The distributions of $V_g$ on $S_h$ with opposite chirality are the same under $s\rightarrow -s$ or $\phi\rightarrow -\phi$.

In the two straight cylindrical leads, the quantum equation (\ref{helieq}) can be separated into longitudinal and angular components analytically. In the case, the wave function with respect to energy $E$ can be written in the form
\begin{equation}
  \psi_{(c)}=\sum_{|m|=0}^{\infty}\chi_m(\phi)(a_m e^{ik_m s}+b_m e^{-ik_m s}),
\end{equation}
where $k_m=\sqrt{2\mu E_s}/\hbar$, $E_s=E-\epsilon_m+\frac{\hbar^2}{8\mu\rho_0^2}$ is the longitudinal energy, $\epsilon_m=\frac{m^2\hbar^2}{2\mu\rho_0^2}$ is the energy of the angular eigenstate $\chi_m(\phi)=\frac{1}{\sqrt{2\pi}}e^{im\phi}$, $a_m$ and $b_m$ are the coefficients for mode $m$ which is the magnetic quantum number with $m=0,\pm1,\pm2,\cdots$. For a given energy $E$, the number of propagating modes ($E_s>0$) in the leads is determined. When $E_s<0$, the wave vector $k_m$ for the corresponding mode $m$ is imaginary, leading to an evanescent (exponentially decaying) state. It should be noted that each two modes with equal absolute value of the quantum number are degenerate, thus the propagating modes always appear in pairs if we raise the total energy $E$.

We assume the electron is injected from $s=0$ and goes out at $s=L$. For an injection state in mode $m$, its wave function is $\psi_{in}=\frac{1}{\sqrt{2\pi}}e^{im\phi}e^{ik_m s}$, the phase term can be written as $im[\phi+(k_m/m)s]$, which is analogous with the term $\phi-\tau s$ mentioned above. Therefore we can also define the chirality of the injection state by the sign of $m$, that is the mode with $m<0$ has the same chirality as the helical tube with $\tau>0$ and vice versa. Now we are able to distinguish the degenerate states by their chiralities. In fact there are many representations to express those degenerate angular states, for instance, the states $(cos(m\phi),sin(m\phi))$. All different forms are related by two dimensional unitary transformations. For example,
\begin{equation} \label{unitary}
\frac{1}{\sqrt{\pi}}\left(
\begin{array}{ccc}
cos(m\phi)\\
sin(m\phi)
\end{array} \right)
=\frac{1}{\sqrt{2}}\left(
\begin{array}{ccc}
1 & 1\\
-i & i
\end{array}\right)\left(
\begin{array}{ccc}
\frac{1}{\sqrt{2\pi}}e^{im\phi}\\
\frac{1}{\sqrt{2\pi}}e^{-im\phi}
\end{array}\right).
\end{equation}
We will refer to the form on the left side of Eq. (\ref{unitary}) as odevity representation, and the form on the right side as chirality representation. In odevity representation, modes with angular states $cos(m\phi)$ and $\sin(m\phi)$ are denoted by subscript $|m|^+$ and $|m|^-$, respectively; and in chirality representation, modes with angular states $e^{im\phi}$ and $e^{-im\phi}$ are denoted by subscript $m$ and $-m$, respectively.

Besides, there are constraints to the geometric parameters for describing a real helical tube in space. An obvious condition is $\rho_0\kappa<1$. For analysis, it is convenient to describe the axis $\textbf{x}(s)$ by the parametrization $(a{\rm cos}\theta^\prime, a{\rm sin}\theta^\prime, c\theta^\prime)$ in Cartesian coordinates, with the curvature $\kappa=\frac{a}{a^2+c^2}$ and torsion $\tau=\frac{c}{a^2+c^2}$. The period length of the line is $l_p=2\pi\sqrt{a^2+c^2}$ corresponding to azimuthal angle $\theta^\prime$ ranging from 0 to $2\pi$. We have to ensure that there is no overlap between $S_h$ at $\theta^\prime=0$ and $\theta^\prime=2\pi$, which means $2\pi c>2\rho_0\frac{\sqrt{a^2+c^2}}{a}$, or equally $\frac{\pi\tau\kappa}{\rho_0(\kappa^2+\tau^2)^{3/2}}>1$. It should also be noticed that, the period of the geometric potential in $s$ direction is $L_p=2\pi/\tau$ , which is longer than $l_p$. The torsion $\tau$ can't be too large to make $l_p$ smaller than the lattice size of nanotubes.

\section{TRANSMISSION SPECTRA}

\subsection{Single propagating mode}
We will first investigate the transport properties of helical nanotubes in the case where only one propagating mode exists, namely the ground state m=0. The transmission of the ground mode in the case of zero torsion (bent cylinders) has been considered in Ref[28]. Here, we adopt the same approach to find the twist effects on the transport property of nanotubes.

\begin{figure}
  \centering
  \includegraphics[width=0.47\textwidth]{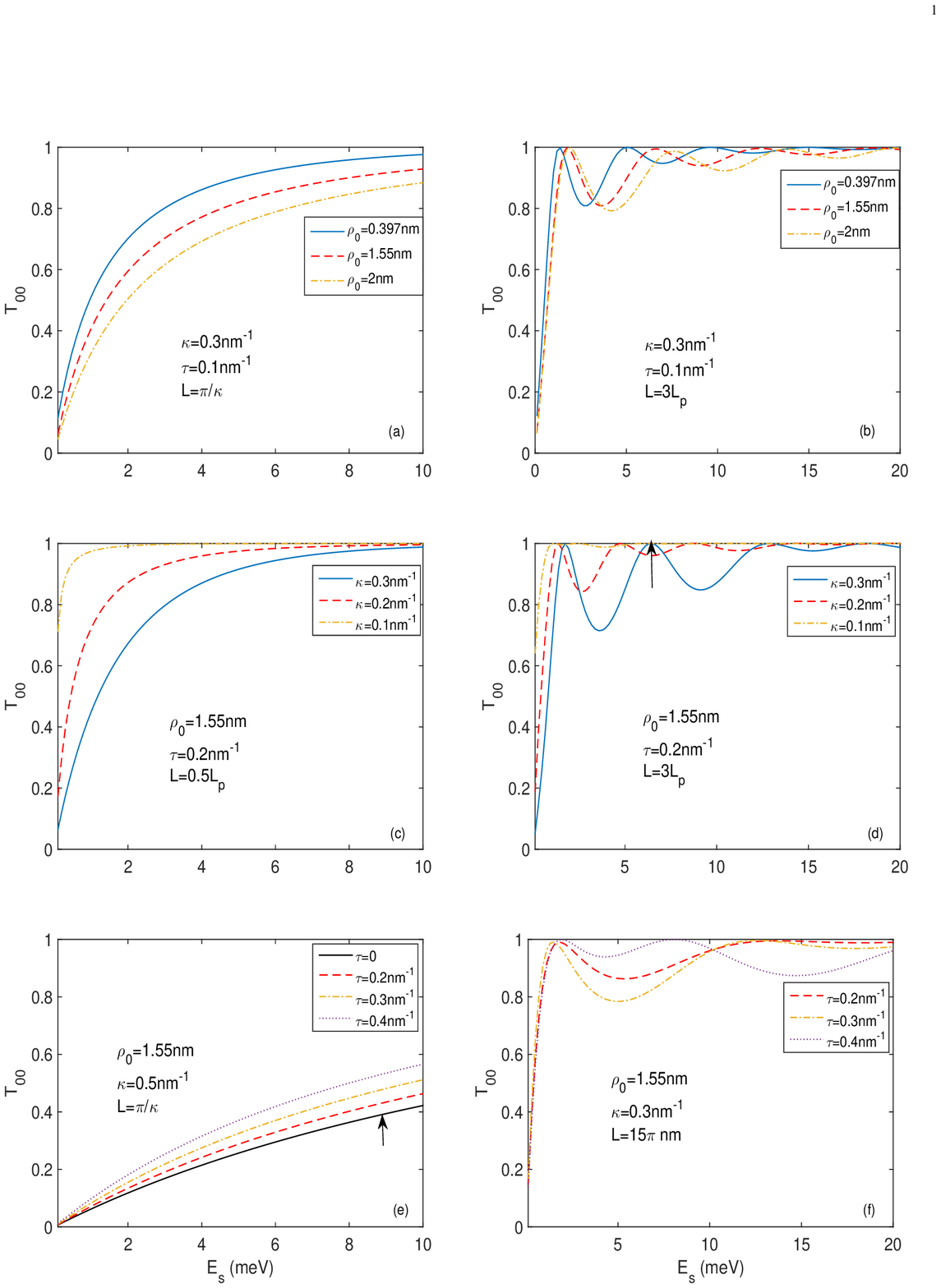}
  \caption{(Color online) Transmission coefficient $T_{00}$ as a function of the injection energy for surfaces of the helical tube with different parameters. The left and right column are for short ($L<L_p$) and relatively long ($L_p<L \leq 3L_p$) $S_h$, respectively. For ground state m=0, the injection energy is $E_s$. }\label{t00}
\end{figure}

For the propagating mode $m=0$, the transmission coefficient as a function of the injection energy $E_s$ is described in Fig. \ref{t00} for the parameters ($\rho_0$, $\kappa$, $\tau$, $L$) with different values, where $L$ denotes the tube length. In Fig. \ref{t00}, $T_{mn}$ is used to indicate the probability that an incident wave in the mode $m$ of one lead is transmitted into the mode $n$ of another lead. For $S_h$ with $L<L_p$, we find $T_{00}$ decreases with increasing $\rho_0$ or $\kappa$, this conclusion is also true in the bent cylindrical case. The increasing leads to the growth of the amplitude of the geometric potential that enhances the reflection component of incident wave. It is obvious that the transmission coefficient $T_{00}$ decreases monotonically with increasing $\tau$. From the right column we find resonant transmission peaks, which correspond to discrete energy levels in $S_h$. It is shown that smaller $\rho_0$ and $\kappa$ induce lower first resonant energies, this is also right for shorter helical nanotubes. The monotonic behaviors of the transmissions for different parameters in the left column, are just related to the position of the first resonant energy. As shown in the left column in Fig. \ref{t00}, the transmissions corresponding to lower first resonant energies grow faster.

It should be noted that the transmission tends to vanish at zero injection energy, which means the reflection probability due to curvature is maximum. In effective mass approximation, the curvature induced geometric potential plays the role of attractive impurities, which cause the incident electron scattering. Carriers with small injection energy will be backscattered by the impurities with high probability, hence a big reflection appears.

\begin{figure}
  \centering
  \includegraphics[width=0.47\textwidth]{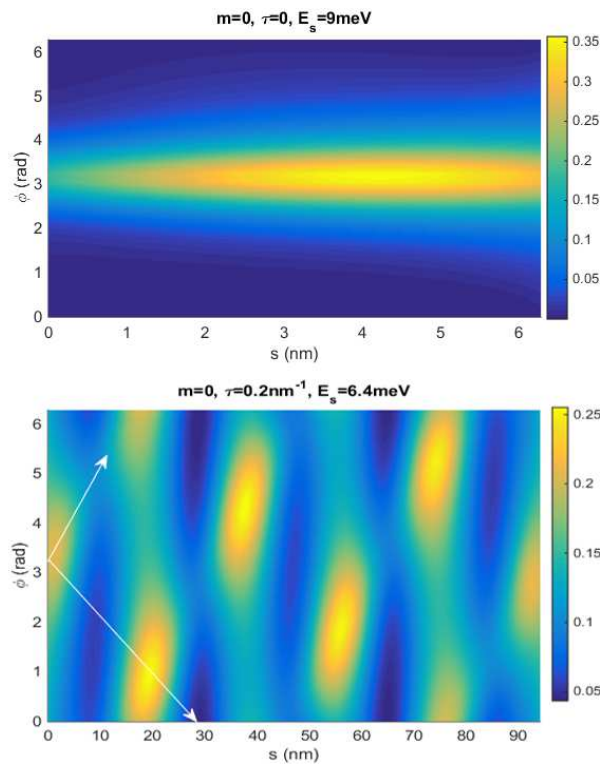}
  \caption{(Color online) The probability density of states in a bent cylindrical (above) and a helical (below) nanotube in the case of an incident wave in mode 0 coming from the boundary s=0. Parameters are corresponding to the solid lines marked by the arrows in Fig. \ref{t00}. The injection energies are $E_s=9$meV (above) and $E_s=6.4$meV (below), respectively. For the helical nanotube, two arrows are used to point the directions along which the maximum probability density could be found. }\label{wave00}
\end{figure}

In the case $L>L_p$, the oscillations of $T_{00}$ indicates that smaller $\rho_0$,$\kappa$ and bigger $\tau$ reduce the intervals between resonant energy levels. Here, the graph for the dependence of $T_{00}$ on $L$ is not shown, however, it's easy to draw a conclusion that longer $S_h$ produces closer energy levels from figures already presented. As there is only one mode $m=0$, the transport can be viewed as a quasi-one-dimensional problem, and we will see from Fig. \ref{wave00}(b) that the carriers tend to propagate along the outer rim of the surface. This is analogous with the simple case of one-dimensional square potential well. Here we utilize the formula from one-dimensional square potential well to explain these oscillations approximately. The effective potential width is $d=L \sqrt{1+\rho^2 \tau^2}$, the potential depth $V=-\frac{\hbar^2}{8\mu \rho^2 (1-\rho \kappa)^2}$ is the minimum of the geometric potential. So the formula for transmission is
\begin{equation} \label{square well}
T_{00}=\left[ 1-\frac{\text{sin}^2 (k^\prime d)}{4 f(1-f)} \right]^{-1},
\end{equation}
where $k\prime=\sqrt{k_0^2+\frac{1}{4\rho^2}[\frac{1}{(1-\rho \kappa)^2}-1]}$, $f=E/V=-(1-\rho \kappa)^2(4\rho^2 k_0^2-1)$. Thus the resonant energy, corresponding to $k^\prime d=n\pi$, is
\begin{equation} \label{resonant energy}
E_s(n)=\frac{\hbar^2}{2\mu}\left\{ \frac{n^2\pi^2}{L^2 (1+\rho^2 \tau^2)}-\frac{1}{4\rho^2}[\frac{1}{(1-\rho \kappa)^2}-1] \right\},
\end{equation}
here, the $n$ is an integer that ensures $E_s(n)>0$, for instance, in Fig. \ref{t00}(d), $n>16$. Eq. (\ref{resonant energy}) shows the dependence of the resonant energy on the parameters in a helical nanotube with low injection energy. The changes in energy levels of the system are all originated from the variation of the shape.

Visualizing the wave functions in $S_h$ can aid in understanding the behaviors of transmission coefficients. In Fig. \ref{wave00} we have plotted the probability density of states in a bent cylindrical and a helical nanotube for incoming wave in mode m=0, at the energy indicated by arrows in Fig. \ref{t00}, respectively. For the bent cylindrical nanotube, as expected, we find that the electrons are tend to localized at the outer rim of the surface where the geometric potential has a minimum. This also occurs in the helical nanotube. The beating patterns are seen along the area where geometric potential valleys appear.
%(see Fig. \ref{potential}).

It is interesting that these patterns can be find along another direction (indicated by the long arrow in Fig. \ref{wave00}), because the distance between the patterns are equal. These patterns are formed by the interference between incoming part and reflected part of the wave. The distance between them is determined by the injection energy, higher energy means shorter wave length, leading to shorter distance. Once the maximum of the probability density appears at the boundary $s=L$, the transmission spectra will have a resonant peak. The injection energy and geometric parameters can determine the position and intensity of the patterns, and therefore, control the behavior of transmission coefficient.
By the way, there are no periodic patterns appearing in the bent cylindrical nanotube because its length is shorter ($L=2\pi$nm) than a wave length. It should be noted that for mode $m=0$, both two pictures are nodeless.

\subsection{Symmetry blocking in a bent cylindrical surface}
If the injection energy $E_i$ reaches the threshold energy $\epsilon_m$, new modes $\pm m$ arise in the leads. Next, we will investigate the transmission spectra for the modes with first excited angular states in two representations, the odevity (denoted by subscripts $1^+$, $1^-$) and chirality (denoted by subscripts $1$, $-1$) representations mentioned above. These two representations are equivalent and mutual referents. We will simply call them $R_o$ and $R_c$ below, respectively.

\begin{figure}
  \centering
 \includegraphics[width=0.47\textwidth]{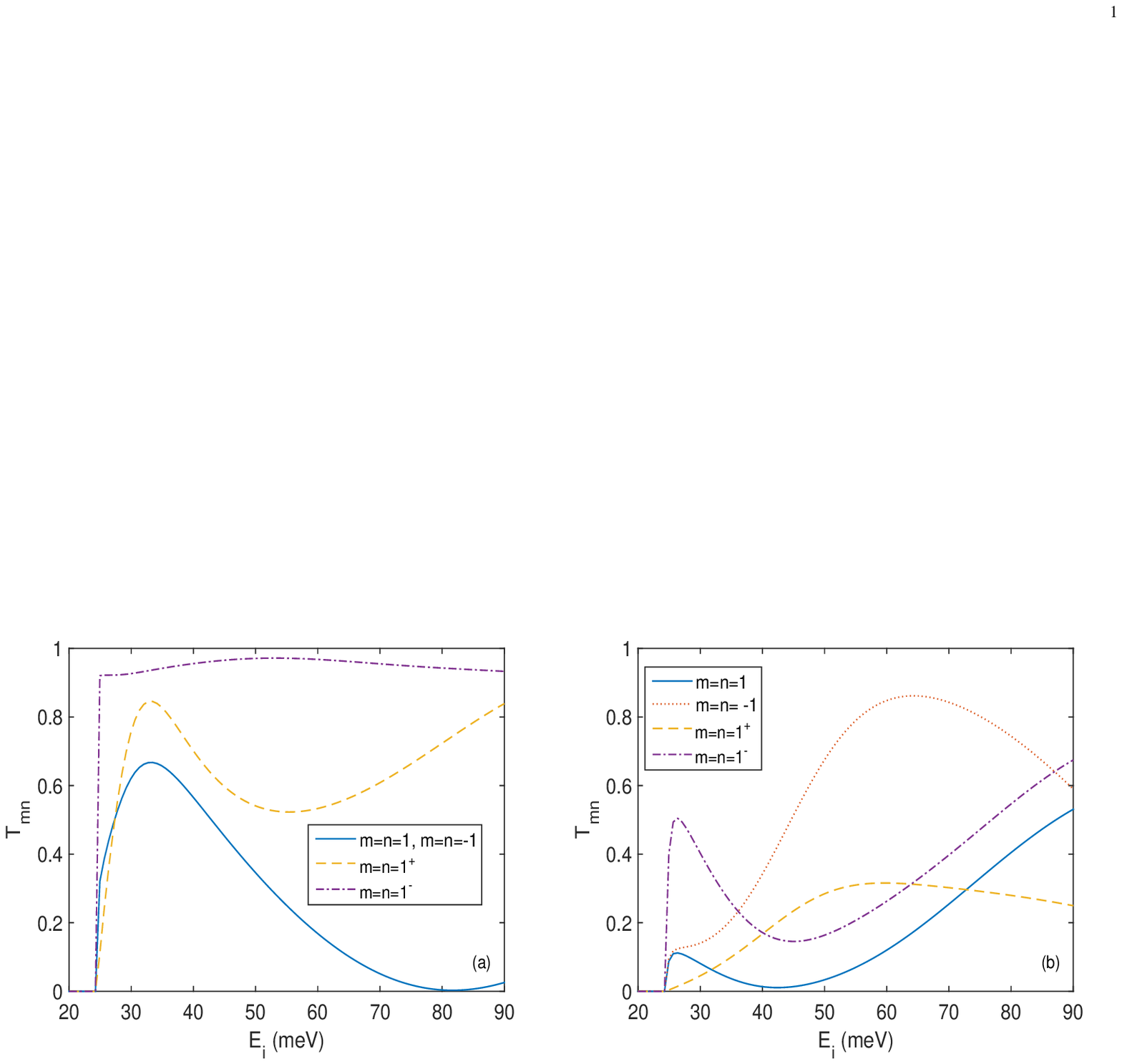}
  \caption{(Color online) Intramode transmission probabilities as functions of injection energy $E_i$ for a bent cylindrical (left) and a helical (right) nanotube with $\tau=0.2$nm$^{-1}$. Calculations are performed in both chirality representation (subscript: $1,-1$) and odevity representation (subscript: $1^+$,$1^-$) for the double-generate first excited states. Common parameters for (a) and (b) are $\rho_0=3$nm, $\kappa$=0.25nm$^{-1}$, $L=\pi/\kappa$. The threshold energy $\epsilon_1=24.47$meV.}\label{t2233}
\end{figure}

A comparison is made in Fig. \ref{t2233} between the intramode transmission probabilities for a bent cylindrical and a helical nanotube. In $R_c$, it is not surprising that for the bent cylindrical one there is $T_{1,1}=T_{-1,-1}$ (here a comma is added between two mode numbers for avoiding ambiguity) due to the symmetry of Hamiltonian $\mathcal{H}(\phi)=\mathcal{H}(-\phi)$ (see Eq. \ref{Hamiltonian} with $\tau=0$). For the helical one, its Hamiltonian has no this symmetry, resulting in $T_{mm}\neq T_{-m-m}$. While in $R_o$, the intramode transmissions are all different for both kinds of nanotubes, owing to the absence of the symmetry $\mathcal{H}(\phi)=\mathcal{H}(\phi \pm \pi/2)$ in each Hamiltonian. In particular we note that once the energy exceeds the threshold $\epsilon_1$, the mode $1^-$ is transmitted almost perfectly.
%This is because the intermode transmissions $T_{1^-, 0}$ and $T_{1^-, 1^+}$ are both too small, which is shown in Fig. \ref{t123}(b).
%This will be explained in the discussion about intermode transmission below.

\begin{figure}
  \centering
\includegraphics[width=0.47\textwidth]{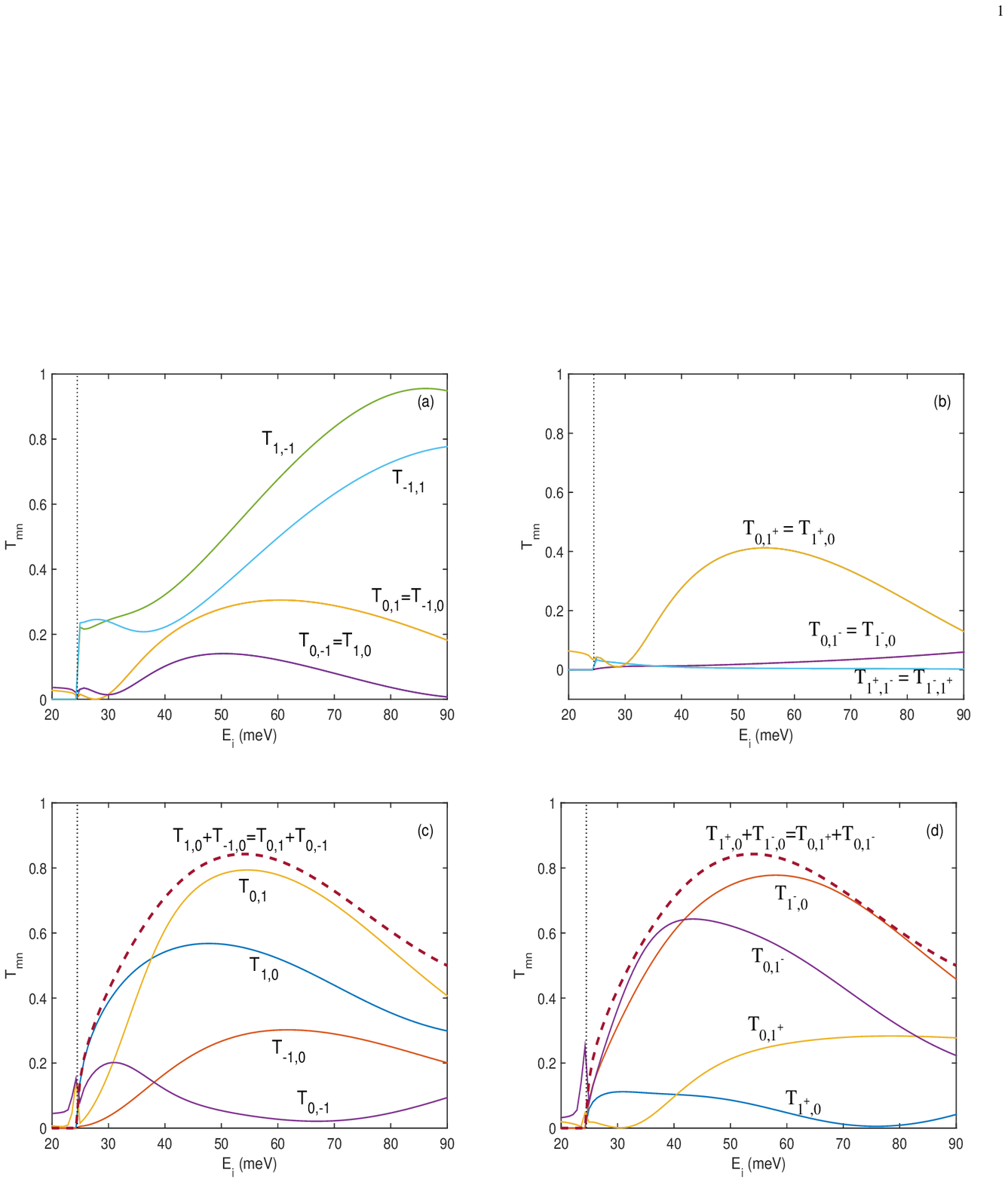}
  \caption{(Color online) Intermode transmission probabilities as functions of injection energy $E_i$ for a bent cylindrical (above) and a helical (below) nanotube with $\tau=0.1$nm$^{-1}$. Calculations are performed in both chirality representation (left, $m=1,0,-1$) and odevity representation (right, $m=1^+,0,1^-$) for the double-generate first excited states. Common parameters are $\rho_0=3$nm, $\kappa$=0.25nm$^{-1}$, $L=\pi/\kappa$. The threshold energy $\epsilon_1=24.47$meV is marked by a vertical dashed line .}\label{t123}
\end{figure}

The intermode transmission probabilities in the two representations for both the bent cylindrical and helical nanotubes are given in Fig. \ref{t123}. Firstly, let us consider the energy range $E_i>\epsilon_1$. It is also noted that for the odd mode $1^-$ in a bent cylindrical nanotube, both $T_{1^-,0}$ and $T_{1^-,1^+}$ are very low in the energy range plotted (see Fig. \ref{t123}(b)). Now it is clear that mode $1^-$ has little probability to convert into another mode, leading to an almost perfect intramode transmission $T_{1^-1^-}$ in Fig. \ref{t2233}(a). The deep reason of this behavior is the symmetry blocking of the geometric potential. This effect is that an incident wave in an odd (even) mode can hardly be converted into an outgoing wave in even (odd) modes in a bent cylindrical surface, as the role of selection rule. Next, we will clarify how the symmetry blocking affects the transport in a bent cylindrical surface.

The Hamiltonian for the bent cylindrical surface $\mathcal{H}_{bc}$ is expressed in Eq. (\ref{Hamiltonian}) with $\tau=0$. Now we will make a rough approximation, and write the Hamiltonian as
\begin{equation} \label{Hbc}
\tilde{\mathcal{H}}_{bc}=-\frac{\hbar^2}{2\mu}\left(\partial_s^2+\frac{\partial_\phi^2}{\rho_0^2}\right)-\frac{\hbar^2}{8\mu\rho_0^2w^2}.
\end{equation}
Here $w=1+\rho_0\kappa{\rm cos}(\phi)$. We remove some kinetic terms from the original Hamiltonian $\mathcal{H}_{bc}$, making the kinetic part of $\tilde{\mathcal{H}}_{bc}$ is the same as it for a straight cylindrical surface (lead). The symmetry $\mathcal{H}_{bc}(\phi)=\mathcal{H}_{bc}(-\phi)$ is still preserved. Then the probability amplitudes for scattering from Lippmann-Schwinger equation[44] can be used,
\begin{equation} \label{LSt}
t_{mn}=\delta_{mn}+\frac{\mu}{i\hbar^2}\sum_{n^\prime}\int ds^\prime \frac{1}{\sqrt{k_n}}e^{-ik_n s^\prime}V_{nn^\prime}\varphi_{n^\prime}^m (s^\prime),
\end{equation}
where $m$ and $n$ denote the mode number of incoming and outgoing wave, respectively,
\begin{equation}
V_{nn^\prime}=-\int d\phi \chi_n^*(\phi)\frac{\hbar^2}{8\mu\rho_0^2w^2}\chi_{n^\prime}(\phi),
\end{equation}
and $\varphi_{n^\prime}^m (s^\prime)$ is from the expansion
\begin{equation} \label{expansion}
\psi_m(s,\phi)=\sum_n\varphi_{n}^m (s)\chi_n (\phi),
\end{equation}
here, $\psi_m(s,\phi)$ is the wave function in the bent cylindrical surface when a wave in mode $m$ is injected.

Because of the symmetry, $\psi(s,\phi)$ must be an even or odd function in $\phi$ direction. In odevity representation, if the incoming wave is in an odd mode, $m=j^-$, $j=1,2,3,...$, as a boundary condition, $\psi(s,\phi)$ is an odd function, similarly, the incoming wave in an even mode corresponds to a wave function which satisfies $\psi(s,\phi)=\psi(s,-\phi)$.
Hence, in the case of incoming wave in an odd mode, all the functions $\varphi_{n}^m(s)$ in Eq. (\ref{expansion}) with $n=j^+$ vanish because of the odevity. For the sum in Eq. (\ref{LSt}), only the terms with $n^\prime=j^-$ are nonzero (here $n=0$ can be viewed as an even mode). Further, as the geometric potential is an even function of $\phi$, if the mode $n$ and $n^\prime$ have different odevities, the matrix element $V_{nn^\prime}$ also vanishes. We have mentioned $n^\prime=j^-$, so only the transmission amplitude with $n=j^-$ possess nonzero value. This means that the incoming wave with an odd mode can only be transmitted into outgoing wave with odd modes. The same analysis for the incident wave in even modes also implies that, the incoming wave with an even mode can only be transmitted into even modes.

Now we know that $T_{1^-,0}$ and $T_{1^-,1^+}$ are small because the mode $1^-$ have different odevity with mode $m=0$ and $1^+$. However, we note that the two transmission coefficients are very low but not zero. These nonzero values are just the contributions of the kinetic terms we have thrown away in Eq. (\ref{Hbc}) from the real Hamiltonian for a bent cylindrical surface.

\subsection{Fano resonance}

\begin{figure}
  \centering
  \includegraphics[width=0.47\textwidth]{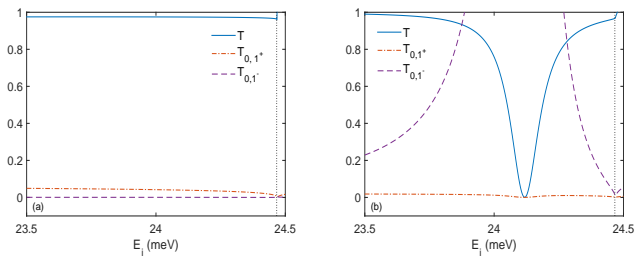}
  \caption{(Color online) Total transmission probability $T$ and intermode transmission coefficients $T_{01^-}$ (dash line) and $T_{01^+}$ (dot-dash line) for a bent cylindrical (left) and a helical (right, $\tau=0.2$nm$^{-1}$) nanotube. Common parameters for (a) and (b) are $\rho_0=3$nm, $\kappa$=0.25nm$^{-1}$, $L=\pi/\kappa$. The threshold energy $\epsilon_1=24.47$meV.}\label{dip}
\end{figure}

In Fig. \ref{t123} we notice that some transmission coefficients have strange behaviors when $E_i$ approaches $\epsilon_1$. For this region, we choose Fig. \ref{t123}(b) and (d) and magnify them in Fig. \ref{dip}, additionally, the total transmission probability $T$ is also presented. By comparing $T$ in both cases, we find that a slightly asymmetric Fano dip going to zero appears at the energy just below the threshold energy $\epsilon_1$ in the transmission spectra for the helical nanotube, however, this doesn't happen in the bent cylindrical nanotube. This dip is due to the quasibound state splitting off from a higher evanescent channel[45]. It is rather clear when we focus on $T_{0,1^-}$ in Fig. \ref{dip}(b), which even exceeds 1 and gets a maximum (which can't be seen in the figure) at the energy of zero point of the dip. It may be surprising that the transmission coefficient is greater than unity, while this is reasonable since the mode $1^-$ is an exponentially decaying evanescent mode when $E_i<\epsilon_1$. The evanescent mode has no contribution to the current in leads, so it is not subject to the current conservation principle. This quasibound state is mainly generated by the bound state of channel $1^-$ getting embedded into the continuum spectrum of channel $m=0$, corresponding to a complex energy. The position of the dip shows the real part of the quasibound state energy, and the imaginary part describes the width of the resonant dip. The channel mixing is mainly due to the geometric potential in $S_h$.

For the bent cylindrical nanotube, both $T_{01^+}$ and $T_{01^-}$ are so small, indicating that no quasibound state causes the Fano resonance. The small $T_{01^-}$ can be interpreted as the symmetry blocking of $V_g$ in a bent cylindrical surface, however this can not explain why $T_{01^+}$ is small around the energy $\epsilon_1$. We guess the reason is that the even (cosinoidal) angular state in the bent cylindrical surface corresponds to a higher energy far away from the threshold energy $\epsilon_1$, hence there is no discrete energy level corresponding to even angular state merges with the continuum spectrum, to generate a quasibound state. Briefly, symmetry blocking effect prevents the forming of quasibound state in bent cylinders, leading to the absence of Fano resonance, while in helical nanotubes, because of the torsion, symmetry blocking does not exist, hence Fano dips appear.

\begin{figure}
  %\centering
 \includegraphics[width=0.47\textwidth]{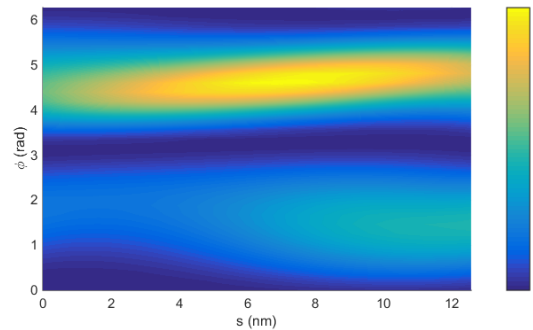}
  \caption{(Color online) The probability density of the state in a helical nanotube in the case of an incident wave in mode $m=0$ coming from the boundary s=0, at the energy corresponding to the minimum of the dip in Fig. \ref{dip}(b). Geometric parameters are the same with Fig. \ref{dip}(b).}\label{wavedip}
\end{figure}

In order to make the physics of the process more clear, in Fig. \ref{wavedip}, we have plotted the probability density of the state in $S_h$ for incoming wave with $m=0$, at the energy corresponding to the minimum of the dip in Fig. \ref{dip}(b). By comparing with Fig. \ref{wave00}, for the energy corresponding to the dip, we find two nodes in the angular direction, $\phi=\pi$ and $\phi=0 (2\pi)$. This nodal structure[44] is in fact the characteristic belong to the next sinusoidal mode $1^-$. The ground mode $m=0$ is resonantly backscattered due to coupling to the evanescent state in the next modes, namely, the mode $1^-$. A beating pattern due to the interference of the incoming and reflected wave is also seen in the region $\phi=3\pi/2$.

\subsection{Symmetry of S-matrix and degenerate state}

In the bent cylindrical case, in $R_c$ we find $T_{0,-1}=T_{1,0}$ and $T_{0,1}=T_{-1,0}$ , and in $R_o$ there is also $T_{m^\pm n^\pm}=T_{n^\pm m^\pm}$, $m,n=0,1$. These equalities are due to the time reversal invariance of the system. After a time reversal, an incoming wave with mode $m$ will become an outgoing wave with mode $-m$ in $R_c$, while in $R_o$, an incoming wave in mode $m^\pm$ will turn into an outgoing wave in the same mode $m^\pm$. So we can deduce that, for a bent cylindrical surface, general relations $T_{mn}=T_{-n,-m}$ and $T_{m^\pm n^\pm}=T_{n^\pm m^\pm}$ are true for any number of propagating modes in $R_c$ and $R_o$ , respectively. This is also the reason why $T_{1,-1}\neq T_{-1,1}$ in Fig. \ref{t123}(a).

However, the two relations are failed in describing the intermode transmissions for a helical nanotube, even though the time reversal symmetry is still preserved. What is more, we find another equality $T_{1,0}+T_{-1,0}=T_{0,1}+T_{0,-1}=T_{1^+,0}+T_{1^-,0}=T_{0,1^+}+T_{0,1^-}$ in this case.

Why cannot the time reversal symmetry ensure the equality of intermode transmission coefficients in the helical nanotube? Here we try to give an explanation from a perspective of S-matrix. The symmetry properties of the scattering and transmission matrices in quantum transport have been precisely deal with in previous literatures[46,47]. Here we consider the problem in coherent transport with double-degenerate injection states. We will show that a pair of double-degenerate states can be viewed as a state with two degrees of freedom, which is reminiscent of fermion spin, thus the violation of the symmetry property of the S matrix can be seen as arising from an effective "magnetic field" induced by the torsion of the nanotube.

Generally, the unitary S-matrix relates the incoming wave $\psi_{in}$ and outgoing wave $\psi_{out}$,
\begin{equation} \label{Soi}
\{\psi_{out}\}=S\{\psi_{in}\},
\end{equation}
where $\{\psi\}$ denotes all possible incoming or outgoing states. It can also be defined in terms of reflection matrices $R$ and transmission matrices $T$[31]
\begin{equation}
S_{mn}=\left[
\begin{array} {ccc}
R_{mn} & T_{mn}^\prime \\
T_{mn} & R_{mn}^\prime
\end{array} \right].
\end{equation}
Because of the absence of magnetic field, the time-reversal symmetry is not broken, we have another relation
\begin{equation} \label{Sio}
\{\psi_{in}\}^*=S\{\psi_{out}\}^*.
\end{equation}
Combining Eq. (\ref{Soi}) and Eq. (\ref{Sio}), and considering the unitary property, we obtain $S^T=S$, which shows the symmetry of S-matrix.

However, if $\{\psi_{in}\}$ and $\{\psi_{out}\}$ contain degenerate states, the problem need to rethink. In our case, degeneration happens in excited angular states in leads. In fact we can not determine the exact form of a pair of double-degenerate states, since any possible form can be transform into another possible form by a unitary matrix $U$. That is, every pair of double-degenerate states corresponds to an arbitrary unitary matrix. For example, in odevity representation, ignoring the longitudinal part, the $m^{th}$ degenerate angular states in incoming wave should be written as
\begin{equation}
\{\psi_{in}\}_m=U\left(
\begin{array} {ccc}
{\rm sin}(m\phi) \\
{\rm cos}(m\phi)
\end{array} \right),
\end{equation}
where $U$ is an arbitrary two dimensional unitary matrix, which can be expressed as
\begin{equation}
U=\left(
\begin{array} {ccc}
a & b \\
-b^* & a^*
\end{array} \right),
\end{equation}
here, $a$ and $b$ satisfy $|a|^2+|b|^2=1$.
Therefore, for degenerate states, in odevity representation, the relation between the $m^{th}$ outgoing degenerate modes and the $n^{th}$ incoming modes is
\begin{equation}
\{U_{out}\}_m\left(
\begin{array} {ccc}
{\rm sin}(m\phi) \\
{\rm cos}(m\phi)
\end{array} \right)=S_{mn}\{U_{in}\}_n\left(
\begin{array} {ccc}
{\rm sin}(n\phi) \\
{\rm cos}(n\phi)
\end{array} \right),
\end{equation}
where $S_{mn}$ is a $2\times 2$ submatrix in $S$. In a fixed representation, both series of matrixes $\{U_{in}\}$ and $\{U_{out}\}$ are determined by initial phases of the incoming wave in $\phi$ direction. So far, it is clear that the transmission coefficients we have calculated for degenerate modes are in fact the elements of
\begin{equation}
\tilde{S}=\{U_{out}^{-1}\}S\{U_{in}\}.
\end{equation}
Once $\{U_{in}\}$ or the initial phases are fixed, $\{U_{out}\}$ are determined automatically. The initial phases are trivial in the transport problem, but the phase differences between the two leads are nontrivial. We assume $\{U_{out}\}=\{W\}\{U_{in}\}$, where $\{W\}$ are also $2 \times 2$ unitary matrices, and set $\{U_{in}\}={I}$, where $I$ is the identity matrix. Following the procedure above, we find $\tilde{S}^T=\{W^{-1}\} \tilde{S} \{W^T\}$. This means the degenerate subspaces experience unitary transformations under the time reversal.

For a bent cylindrical nanotube, the wave function can be separated $\psi(s,\phi)$=$\psi(s)\psi(\phi)$, so the phase in $\phi$ direction at $s=0$ are identical to it at $s=L$, which means the initial phase is preserved during the transmission. Thus, we can deduce $\{U_{out}\}= \{U_{in}\}$. This equality results in $\tilde{S}_{mn}=\tilde{S}_{nm}$, which is observed in Fig. \ref{t123}. For the helical nanotube, because of the coupling phase $\phi-\tau s$, the wave function cannot be separated in two parts, showing no property to keep the initial phase along $s$ direction, the unitary matrices $\{W\}$ are no longer identity matrices. In this case, $\tilde{S}$ loses the symmetry. However, due to the unitarity of $U_{in}$ and $U_{out}$ in $\tilde{S}$ and time reversal symmetry of the system, we can still prove that, in any representation, there is
\begin{equation} \label{S}
\sum_{i,j} S_{m_i,n_j}=\sum_{i,j} S_{n_i,m_j},  \qquad i,j=1,2,
\end{equation}
where $m_1$ and $m_2$ denote two components of the $m^{th}$ double-degenerate states in a specific representation, respectively. The equalities $T_{1,0}+T_{-1,0}=T_{0,1}+T_{0,-1}=T_{1^+,0}+T_{1^-,0}=T_{0,1^+}+T_{0,1^-}$, which are showed in Fig. \ref{t123}, are special examples of Eq. (\ref{S}). Eq. (\ref{S}) also implies that the total transmission is independent of the choice of representations.

Therefore, from an overall perspective, the double-degenerate states are equal to a state with two degrees of freedom, similar to the spin of a fermion. For a spin-1/2 system, a magnetic field can lift the degeneracy of spin, giving asymmetric properties for particles with opposite spin orientations, and break the time reversal symmetry of the system. In our case, the coherent transport is also asymmetric for the two components of degenerate states, and the time reversal symmetry is partly broken in the subspaces form by degenerate states. So the torsion of the nanotube may act as an effective magnetic field which induces a "Zeeman coupling" between the "magnetic field" and the degenerate subspaces.

\section{CONDUCTANCE}
\begin{figure}
  \centering
  \includegraphics[width=0.47\textwidth]{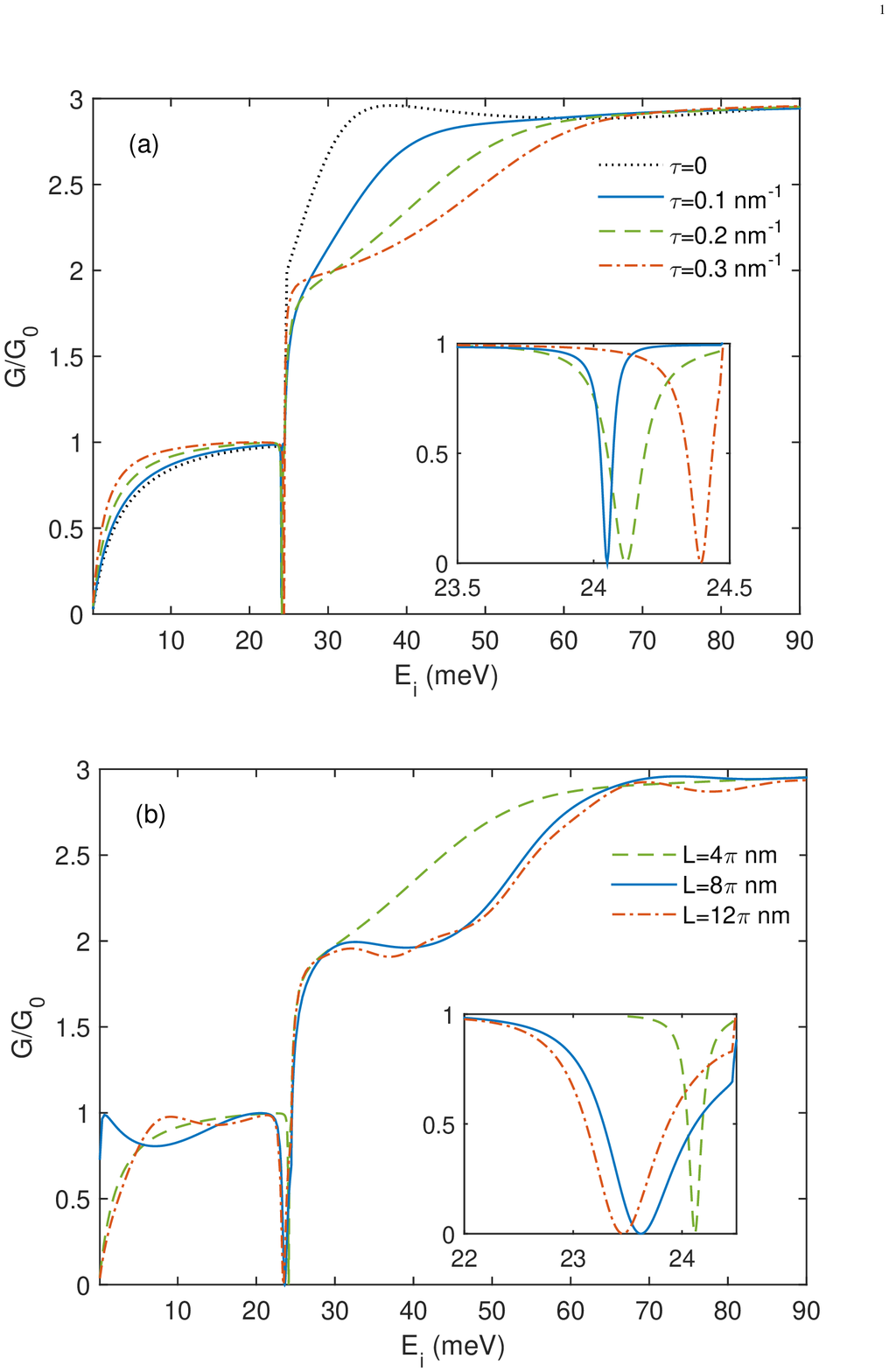}
  \caption{(Color online) Conductance of a helical nanotube in units of $G_0=2e^2/h$ as a function of injection energy at zero temperature, for different $\tau$ (upper panel) and different length $L$ (lower panel). Other geometric parameters: (a) $\rho_0=3$nm, $\kappa$=0.25nm$^{-1}$, $L=\pi/\kappa$; (b) $\rho_0=3$nm, $\kappa$=0.25nm$^{-1}$, $\tau=0.2$nm$^{-1}$. Note that the conductance for a bent cylindrical nanotube is also presented in (a). A detailed view around the dips is presented in the insets. The threshold energy $\epsilon_1=24.47$meV.}\label{con}
\end{figure}

Based on Landauer conductance formula[48,49], the conductance of helical nanotubes versus the injection energy $E_i$ at zero temperature is shown in Fig.~\ref{con}, for different $\tau$ and $L$. The bias is assumed very small. Apparent deviations from the steplike structure belong to straight cylindrical surfaces appear. For a straight cylinder, the double-degenerate modes contribute equally to the conductance[50], an abrupt step from $G_0$ to $3G_0$ should arise at the threshold energy $\epsilon_1$, where $G_0=\frac{2e^2}{h}$ is the quantum of conductance, with $h$ the Planck's constant. The deviations show the geometric effect on conductance.

In a short ($L=\pi/\kappa$) helical nanotube, the conductances immediately reach about $2G_0$ after the threshold, and then get to $3G_0$ through an approximate linear growth. By increasing the torsion $\tau$, the growth from $2G_0$ to $3G_0$ becomes slower, leading to shorter plateau of $3G_0$. While if the nanotube becomes longer, without changing other parameters, we find the growth from $2G_0$ to $3G_0$ is no longer linear, instead, a new plateau of $2G_0$ with width about 20meV is formed. It seems that one of the double-generate states propagates perfectly, while the other one is completely reflected in the energy range of this plateau. In the helical nanotube, eigenstates are no longer degenerate, leading to the separation of transmission channels for a pair of degenerate modes. We have conclude that longer nanotubes create closer energy levels. Therefore, the more concentrated energy levels exacerbate this separation and give a new plateau.

Next we focus on the variation of the dip induced by changes of $\tau$ and $L$. As mentioned above, this Fano resonance is caused by a bound state being embedded in the continuum of the energy spectrum, forming a quasibound state. Both $\tau$ and $L$ can change the energy spectrum, including the energy of the bound state and continuum, and then change the dips. From insets of Fig. \ref{con}, the dip energy shifts to right with $\tau$ increasing, but with $L$ decreasing. So we conclude that the real part of the quasibound state energy in a helical nanotube is increased by the twist, and decreased by its length. In a longer nanotube, the dip is clearly wider than it's in a shorter one, which means the quasibound state energy getting a greater imaginary part. We stress that the length of the longer nanotube is smaller than $2L_p$ here. For a long helical nanotube, which is beyond the scope of this article, the periodicity of geometric potential will make the transport property more complicated, thus the conclusion we have mentioned above may be not suitable. In addition, the conductance for a nanotube with smaller $\rho_0$ will possess wider plateaus because the threshold energy $\epsilon_m$ is inversely proportional to $\rho_0^2$.

\section{Conclusion}
In this work, the expected Schr\"{o}dinger equation for a particle bound to the surface of a helical nanotube has been briefly given by the thin-layer quantization procedure. The geometric potential in the equation is periodic in both $\phi$ and $s$ directions, but is a constant in the direction $\phi=\tau s$.

Using QTBM, we have calculated the transmission probability for the injected wave in the single mode and multimode. In the single propagating mode case, before the first resonant transmission peak appears, the transmission coefficient increases monotonically with increasing $\rho_0$ and $\kappa$, but decreases with increasing $\tau$. The energy intervals are substantially affected by $\rho_0$, $\kappa$, $\tau$ and $L$. For the probability density of state in the helical nanotube, periodic beating patterns appear along the outer rim of the helical nanotube, showing nodeless characteristic. In multimode case, Fano resonance appears as a multimode phenomenon purely originated from the geometric potential in helical nanotubes, but doesn't happen in bent cylindrical systems. This resonance is due to quasibound state splitting off from the evanescent channel. Symmetry blocking in bent cylindrical surface, owing to the symmetry in angular direction, has been discovered and interpreted in Lippmann-Schwinger equation. The absence of Fano dip in the conductance for bent cylindrical nanotubes can be explained by this effect.

Equalities on transmission coefficients induced by the symmetries of the systems are found. By distinguishing the double-degenerate states in a certain representation, we clarify that the transmission coefficients are the elements of a modified S-matrix. It is found that for the bent cylindrical system and the helical system, this matrix has the different symmetry. The property of the modified S-matrix shows that the torsion of the helical nanotube can induce unitary transformations to the degenerate subspaces under time reversal.

Finally, the conductance of helical nanotubes for different torsion and length was analyzed. It is found that the two parameters can significantly change the behaviors of the conductance, especially shift and change Fano dip. A new plateau appears in the conductance when the length of the nanotube is suitable.

\acknowledgments

This work is supported by the National Natural Science Foundation of China (under Grants No. 11274166, No. 11275097, No. 11475085, and No. 11535005).

%\bibliographystyle{apsrev4-1}
%\bibliography{reference_liang}

\begin{thebibliography}{99}

\bibitem{1} {G. Cantele, D. Ninno, G. Iadonisi, J. Phys.: Condens. Matter 12} (2000) {9019.}
\bibitem{2} {R. Dandoloff, T. Truong, Phys. Lett. A 325} (2004) {233.}
\bibitem{3} {V.M. Fomin, S. Kiravittaya, O.G. Schmidt, Phys. Rev. B 86} (2012) {195421.}
\bibitem{4} {A. Bachtold, C. Strunk, J.-P. Salvetat, J.-M. Bonard, L. Forr車, T. Nussbaumer,
C. Sch?nenberger, Nature 397} (1999) {673.}
\bibitem{5} {Y. Zhang, G. Yu, J. Dong, Phys. Rev. B 73} (2006) {205419.}
\bibitem{6} {C.-H. Chang, J. van den Brink, C. Ortix, Phys. Rev. Lett. 113} (2014) {227205.}
\bibitem{7} {H. Jensen, H. Koppe, Ann. Phys. 63} (1971) {586.}
\bibitem{8} {R.C.T. da Costa, Phys. Rev. A 23} (1981) {1982.}
\bibitem{9} {Y.-L. Wang, H.-S. Zong, Ann. Phys. 364} (2016) {68.}
\bibitem{10} {C. Ortix, J. van den Brink, Phys. Rev. B 81} (2010) {165419.}
\bibitem{11} {Z. Li, L.R. Ram-Mohan, Phys. Rev. B 85} (2012) {195438.}
\bibitem{12} {V. Atanasov, R. Dandoloff, A. Saxena, Phys. Rev. B 79} (2009) {033404.}
\bibitem{13} {I. Blonskyy, V. Kadan, A. Kadashchuk, A. Vakhnin, A. Zhugayevych, I. Chervak, Phys. Low-Dimens. Struct. 7每8} (2003) {25.}
\bibitem{14} {O. Bisi, S. Ossicini, L. Pavesi, Surf. Sci. Rep. 38} (2000) {1.}
\bibitem{15} {J. Onoe, T. Ito, H. Shima, H. Yoshioka, S. ichi Kimura, Europhys.
Lett. 98} (2012) {27001.}
\bibitem{16} {J. Kim, V. Chua, G.A. Fiete, H. Nam, A.H. MacDonald, C.-K. Shih, Nat.
Phys. 8} (2012) {464.}
\bibitem{17} {A. Szameit, F. Dreisow, M. Heinrich, R. Keil, S. Nolte, A. T邦nnermann, S.
Longhi, Phys. Rev. Lett. 104} (2010) {150403.}
\bibitem{18} {R. Saito, G. Dresselhaus, M.S. Dresselhaus, Phys. Rev. B 53} (1996) {2044.}
\bibitem{19} {R. Tamura, M. Tsukada, Phys. Rev. B 61} (2000) {8548.}
\bibitem{20} {W. Lu, E.G. Wang, H. Guo, Phys. Rev. B 68} (2003) {075407.}
\bibitem{21} {T. Ando, J. Phys. Soc. Japan 74} (2005) {777.}
\bibitem{22} {W. Lu, Sci. Technol. Adv. Mater. 6} (2005) {809.}
\bibitem{23} {J. Geng, K. Kim, J. Zhang, A. Escalada, R. Tunuguntla, L.R. Comolli, F.I.
Allen, A.V. Shnyrova, K.R. Cho, D. Munoz, et al., Nature 514} (2014) {612.}
\bibitem{24} {E.A. Laird, F. Kuemmeth, G.A. Steele, K. Grove-Rasmussen, J. Nyg?rd, K.
Flensberg, L.P. Kouwenhoven, Rev. Mod. Phys. 87} (2015) {703.}
\bibitem{25} {C.-H. Kiang, M. Endo, P.M. Ajayan, G. Dresselhaus, M.S. Dresselhaus, Phys.
Rev. Lett. 81} (1998) {1869.}
\bibitem{26} {W. Guo, W. Zhong, Y. Dai, S. Li, Phys. Rev. B 72} (2005) {075409.}
\bibitem{27} {C. Tang, W. Guo, C. Chen, Phys. Rev. B 79} (2009) {155436.}
\bibitem{28} {A. Marchi, S. Reggiani, M. Rudan, A. Bertoni, Phys. Rev. B 72} (2005) {035403.}
\bibitem{29} {L. Du, Y.-L. Wang, G.-H. Liang, G.-Z. Kang, X.-J. Liu, H.-S. Zong, Physica
E 76} (2016) {28.}
\bibitem{30} {J. Chen, L. Li, D. Zhou, X. Wang, G. Xue, Phys. Rev. E 92} (2015) {032306.}
\bibitem{31} {J. Goldstone, R.L. Jaffe, Phys. Rev. B 45} (1992) {14100.}
\bibitem{32} {G. Cuoghi, A. Bertoni, A. Sacchetti, Phys. Rev. B 83} (2011) {245439.}
\bibitem{33} {W. Yan, W.-Y. He, Z.-D. Chu, M. Liu, L. Meng, R.-F. Dou, Y. Zhang, Z. Liu,
J.-C. Nie, L. He, Nat. Commun. 4} (2013)
\bibitem{34} {H. Taira, H. Shima, J. Phys. Condens. Matt. 22} (2010) {075301.}
\bibitem{35} {C. Ortix, Phys. Rev. B 91} (2015) {245412.}
\bibitem{36} {A. Javey, J. Guo, M. Paulsson, Q. Wang, D. Phys. Rev. Lett. 92} (2004) {106804.}
\bibitem{37} {D. Mann, A. Javey, J. Kong, Q. Wang, H. Dai, Nano Lett. 3} (2003) {1541.}
\bibitem{38} {C. Berger, Y. Yi, Z. Wang, W. de Heer, Appl. Phys. A 74} (2002) {363.}
\bibitem{39} {L.C. Venema, J.W. Janssen, M.R. Buitelaar, J.W.G. Wild?er, S.G. Lemay, L.P.
Kouwenhoven, C. Dekker, Phys. Rev. B 62} (2000) {5238.}
\bibitem{40} {A. Rochefort, D.R. Salahub, P. Avouris, Chem. Phys. Lett. 297} (1998) {45.}
\bibitem{41} {C.S. Lent, D.J. Kirkner, J. Appl. Phys. 67} (1990) {6353.}
\bibitem{42} {P. Ordej車n, E. Artacho, J.M. Soler, Phys. Rev. B 53} (1996) {R10441.}
\bibitem{43} {D. S芍nchez-Portal, P. Ordej車n, E. Artacho, J.M. Soler, Int. J. Quant.
Chem. 65} (1997) {453.}
\bibitem{44} {J.H. Bardarson, I. Magnusdottir, G. Gudmundsdottir, C.-S. Tang, A. Manolescu,
V. Gudmundsson, Phys. Rev. B 70} (2004) {245308.}
\bibitem{45} {P.N. Racec, E.R. Racec, H. Neidhardt, Phys. Rev. B 79} (2009) {155305.}
\bibitem{46} {M. Buttiker, IBM J. Res. Dev. 32} (1988) {317.}
\bibitem{47} {M. B邦ttiker, Phys. Rev. B 38} (1988) {9375.}
\bibitem{48} {R. Landauer, IBM J. Res. Dev. 1} (1957) {223.}
\bibitem{49} {R. Laudauer, Phil. Mag. 21} (1970) {863.}
\bibitem{50} {J.A. Torres, J.I. Pascual, J.J. S芍enz, Phys. Rev. B 49} (1994) {16581.}


\end{thebibliography}

\end{document}